\documentclass{iopart}
\usepackage{iopams}
\usepackage[latin1]{inputenc}
\usepackage[T1]{fontenc}
\usepackage[english]{babel}
\usepackage{cite} 
\usepackage{textcomp}
\usepackage{graphics}
\usepackage{graphicx}
\usepackage{anysize}
\usepackage{amssymb}
\usepackage{epsfig}
\usepackage{subfigure}
\usepackage{float}
\usepackage{wrapfig}

\begin{document}
\title{The mechanism of long-term coarsening of granular mixtures in rotating drums}
\author{Tilo Finger$^1$, Matthias Schr\"oter$^2$, Ralf Stannarius$^1$}
\address{$^1$Institute of Experimental Physics, Otto-von-Guericke-University, Universit\"atsplatz 2, D-39106 Magdeburg, Germany, $^2$Max Planck Institute for Dynamics and Self-Organization (MPIDS), 37077 Goettingen, Germany}

\ead{tilo.finger@gmx.de, matthias.schroeter@ds.mpg.de}
\begin{abstract}
Three fundamental segregation and pattern formation processes are known in granular mixtures in a rotating
cylindrical drum: radial segregation, axial banding, and coarsening of the band pattern. While the mechanism
for the first effect is well understood and for the second effect, several models have been proposed,
the long-term coarsening mechanism remained unexplained so far.
We demonstrate that the unidirectional flow between the bands in an axially segregated pattern is driven by small
differences in size of the small beads at the band edges.
Due to a process of microsegregation inside each band of small particles, which was so far unrecognized,
this difference in diameter will be effective in all experiments with polydisperse beads. In consequence
the stability of individual bands can be easily controlled by minor alterations of their composition.
Our results make evident that a new mechanism as the driving force behind the axial particle flow has to be sought.
We suggest possible hypotheses for such a mechanism.
\end{abstract}

\section{Introduction}
Granular mixtures in a rotating cylindrical drum represent one of the fundamental dynamic experiments in granular matter research.
First described already in 1939 \cite{Oyama39,Oyama40}, pattern formation and dynamics in this system
remained attractive scientific topics until now, both in experiment and theory
\cite{Zik94,Nakagawa94,Alexander04,Fiedor03,meier:07,schlick:15,Inagaki10,Gupta91,aranson:99,aranson:99_PRE,Hill94,Hill95,Clement95,Cantelaube95,Hill97,Hill97a,Nakagawa97,Ristow99,
Chen10,Chen11,Taberlet04,Khan05,Taberlet06,zuriguel:05,pohlman:06,Fischer09,Kuo06,Arndt05,Finger06,Finger07,Juarez08,
Frette97,Choo97,Choo98,khan:11,nguyen:11,rapaport:02,rapaport:07}.
Features of interest range, e.g., from the variation of particle parameters
\cite{Alexander04} and drum geometries \cite{Fiedor03,meier:07,schlick:15}, filling aspects \cite{Inagaki10}, and segregation mechanisms
\cite{Gupta91,Clement95,Cantelaube95,Hill97,Hill97a,Nakagawa97,Hill94,Hill95,Ristow99,Chen10,Chen11,aranson:99,aranson:99_PRE} to
particle \cite{Taberlet04,Taberlet06,Khan05,pohlman:06,Fischer09,zuriguel:05} and pattern dynamics \cite{Choo97,Choo98,Inagaki10,Kuo06,Frette97,Juarez08,Arndt05,Finger06,Finger07}.
A recent review that gives a certain overview over studies related to the rotating granular mixer can be found
in Ref. \cite{Seiden11}.

Apart from the fundamental scientific interest in the dynamical features of this simple multiparticle system,
impact on technological processes cannot be overestimated. In most applications, segregation of mixtures of grains is
unwanted, in some others it may be highly desired. General understanding of the underlying physical mechanisms is
therefore of considerable practical interest.

Segregation of a binary granular mixture in a rotating drum typically happens in three stages. First, {\it radial  segregation}  leads
to the formation of a core region consisting of the smaller particles \cite{Clement95,Hill97,Nakagawa97,nguyen:11}. The emergence of the core is well understood as the consequence of a kinetic sieving effect. However, there is still an open scientific debate \cite{Taberlet06, Fischer09, Third11, Third12, Seiden11, khan:11,Christov12}, triggered by Khan and Morris \cite{Khan05}, whether the transport of the beads along the core is subdiffusive or diffusive.

In many systems radial segregation is followed by {\it axial segregation}, the formation of a banded pattern
\cite{Gupta91,Hill94,Hill95,aranson:99,aranson:99_PRE,Taberlet04,taberlet:06_b,Chen10,Chen11,rapaport:07}.
Several models have been proposed for this effect, and solid experimental evidence has been collected to evaluate
proposed mechanisms (see, e.g. Ref. \cite{Alexander04} and references therein).

Finally, {\it coarsening} of  this pattern
\cite{Fiedor03,Nakagawa94,Frette97,Juarez08,Arndt05,Finger06,Finger07} results in
a logarithmic decay of the number of axial bands  with the number of revolutions.
This long-term dynamics happens on the timescale of thousands of rotations, it is clearly
distinguished from a topologically similar merging of bands during the very early phase of axial segregation \cite{Taberlet04,taberlet:06_b}.
While the process of coarsening can be clearly attributed to the exchange of small beads between neighboring bands
\cite{Hill97,Arndt05,pohlman:06,sanfratello:09,nguyen:11}, the mechanism driving this
flux has not been identified so far. It cannot be explained intuitively, nor does it seem to have
relations to any other known dynamic segregation phenomena in granular matter.

\begin{figure}[t]
\centerline{  \includegraphics[width=0.6\textwidth]{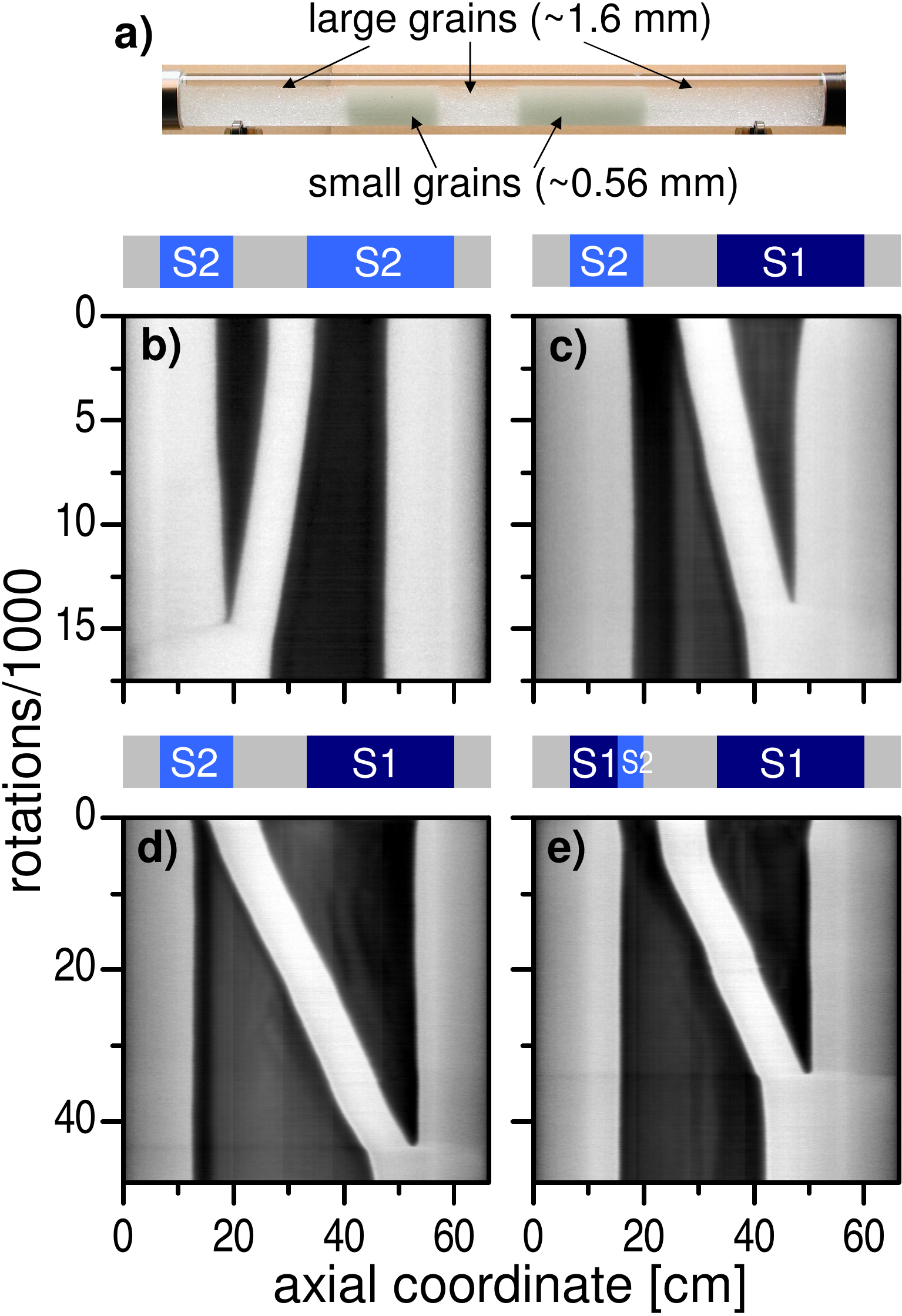}}
  \caption{The flux of small particles between two stripes is determined by both
      the stripe width ratio and the size of particles at the opposing stripe edges.
     {\bf a)} image of a mixer with two prepared stripes of small particles embedded in large-size beads.
      Panels {\bf b-e} show space-time plots of four scenarios.
      They are generated by stacking horizontal lines taken from subsequent images such as panel a).
      The central stripe of large beads (bright) is always 8~cm broad and separates two (darker) stripes
      of (S1) or (S2) particles (sizes see text).
      Top bars sketch the preparation conditions.
      {\bf b)} For same grain types, here S2, the narrower stripe (10.5 cm versus 14 cm) dissolves.
      {\bf c)} If a stripe is prepared from the larger grains S2 (same widths as in b),
      the one with S1 grains dissolves.
      Panel {\bf d)} demonstrates the dominance of grain diameter over stripe width (4~cm S2 versus 30~cm S1).
      {\bf e)} Even a small amount (0.5~cm) of S2 beads on the
      inner edge of the left stripe with 7.5 cm S1 beads,
      triggers dissolution of the broad stripe (20 cm) with purely S1 beads.
      }
  \label{Fig1}
\end{figure}

The elementary process of this exchange was studied optically and with NMR imaging \cite{Finger06}: The mixer was initially
prepared with two separate bands of small-sized beads, enclosed by bands of the large-sized grain component (cf. Fig. \ref{Fig1}a). It was observed that the transport between two adjacent bands is unidirectional. The narrower band loses material to the larger one
at a constant rate, until it is extinguished \cite{Finger06,Finger07}. Once this transport sets in, its direction is irreversible.

In contrast, the large beads do not contribute to the coarsening dynamics. The reason for this is that large grains cannot pass neighboring bands of small grains which are free of large beads \cite{Finger06}. When a band of small beads separating two large-grained bands is extinguished, the latter simply merge to one region.

If the experiment is continued for sufficiently long
time (up to weeks), the result is complete segregation
into one or two bands of large grains and another band plus the core channel containing the smaller grains.

This feature, albeit described in numerous papers, is unexplained so far. Particularly, the following aspect is remarkable: The agitation
of the grains occurs by revolutions about the cylinder axis, while the directional transport is axial. It breaks the axial symmetry of the
 horizontal tube.
Consequently one has to search for a mechanism that drives a unidirectional flow of the small grains, preferentially from smaller to
broader segregation bands.
The process reminds of a hydrodynamic situation where one container empties into a communicating vessel by pressure differences.
Thus, one potential hypothesis for the driving mechanism of the flux could be an effective surface tension at the boundery of large and small  grains. Such a property could originate from an decrease in a configurational
entropy \cite{pica-ciamarra:12,bi:15} at the interface, as discussed below in Sec.~\ref{sec:hypo}.

In this study, we demonstrate that two cooperative processes are responsible for the coarsening.
First, we show that if two adjacent stripes of small beads contain grains of slightly different sizes near their edges,
the transport of grains through the core channel is always from the edge containing smaller grains towards the edge containing
larger grains\footnote{For a visualisation of a core channel see: https://www.youtube.com/watch?v=dN0145p38dU.  This rendering of
a tomographic image corresponds to the experiment discussed in figure 4.}. Secondly, we reveal a microsegregation mechanism inside the band of small beads:
When the size distribution of the small beads is not
exactly monodisperse, larger ones accumulate at the borders of the segregation stripes.
The center of the stripes contains smaller grains.

These results have three important consequences: (1) they provide an explanation why directed transport proceeds from narrower
to broader neighboring stripes containing initially the same grain compositions,
(2) they allow to control the stability of individual stripes
by adding a small amount of slightly larger grains, and
(3) they lead to the prediction that bands of monodisperse small grains
do not show merging or coarsening. We verify this prediction experimentally.

\section{Experimental setup and materials}

The experiments are performed in a 66~cm long tube of 37~mm diameter, rotated at 20 revolutions per minute (not critical).
The tubes are half
filled with bands composed of large (1.62 $\pm$ 0.062 mm diameter in all experiments)
and smaller glass spheres. The latter are obtained by sieving mixtures, we use two fractions with
size distributions between 0.355 mm and 0.500 mm ('S1')
and between 0.500 mm and 0.630 mm ('S2')~\footnote{One can, in principle, consider this a ternary mixture, but since the two neighboring sieving fractions combined form a practically continuous size distribution, this system is qualitatively different from the classical ternary systems described in literature.}. The tube is filled up with water after preparation of the stripes (Fig. \ref{Fig1}a).
The interstitial fluid has only quantitative but no qualitative consequences for the
 coarsening process \cite{Finger07}.
We use water primarily to avoid static charging and to improve sample transparency. The tube is illuminated from the
top and images are recorded automatically in intervals of 1~min.

From long-term observations (between 15,000 and 200,000 rotations), we construct space-time plots using a horizontal line from each image and stacking them together in vertical direction. Regions with small beads S1, S2 appear darker than those of large beads.

\section{Experimental results}

\subsection{Merging of stripes}

\begin{figure}[htbp]
	\begin{center}
	 \includegraphics[width=0.8\textwidth]{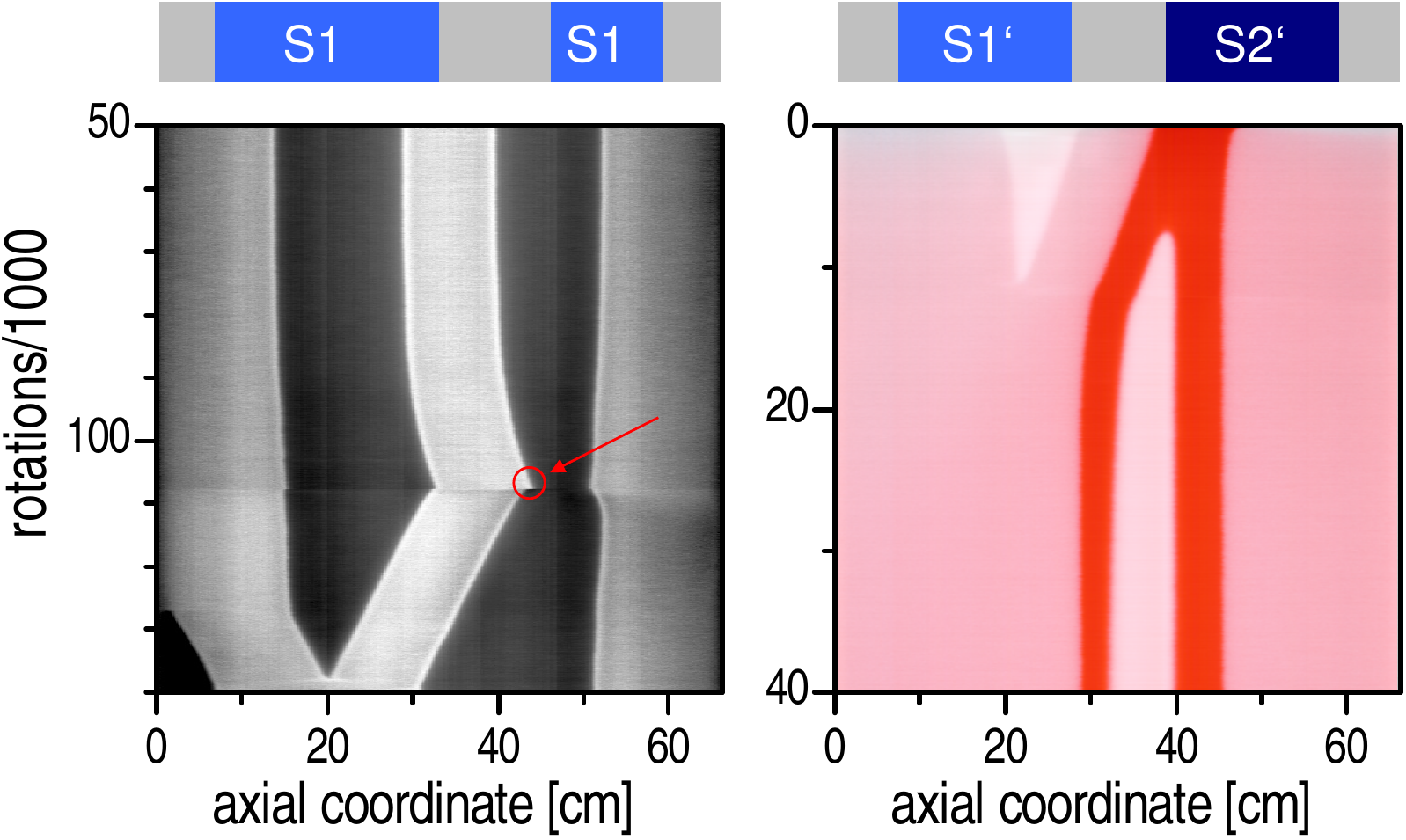}
	\end{center}
	\caption{Coarsening control and microsegregation.
{\bf Left:} After the narrower of two prepared pure S1 stripes starts to dissolve, we add an small amount S2 material (equivalent to 1 cm band)
 to the dissolving 6.5 cm broad S1 band (circled region). This reverses the transport and the 17.5 cm broad S1 stripe starts to decay immediately.
{\bf Right:}  Stripes of smaller white (S1$'$) and larger red beads (S2$'$, see text) beads were prepared. A core quickly forms from material of both stripes
(thus, the regions of large grains also adopt a pink appearance). The S1$'$ band transfers its material to the S2$'$ band,
where it is not simply deposited at the edge but forms a clearly segregated central band, surrounded by S2$'$ beads.
\label{Fig3}}
\end{figure}

Figure \ref{Fig1} b-e shows four typical scenarios with differently prepared initial configurations. The plots demonstrate that
(1) if two bands consist of the same particle types, the narrower one vanishes, see Fig. \ref{Fig1}b,
(2) if one band contains slightly smaller grains than the other, it vanishes even if it is broader,
see Figs. \ref{Fig1}c,d, and (3) it is sufficient that the band edge contains larger particles to stabilize it,
see Fig. \ref{Fig1}e. This allows us to stabilize a dissolving band by adding a few S2 grains
(Fig.~\ref{Fig3}, left):
After a narrow band of S1 grains had already lost half of its content to the neighboring S1 band,
we added $\approx$ 15~\% of S2 material to it\footnote{The water in the tube is drained without perturbing the granular layer. Then a small amount of grains is added on top of the layer with a long stemmed "spoon". Finally, the tube is filled again with water. After approximately 50 rotations, the added material is completely absorbed in the granular bed.}. This led to the immediate reversal of the transport through the core channel, even though the competing band contained almost 3 times as much material.

\subsection{Microsegregation}

The second process active in the coarsening is microsegregation inside each individual band of small particles.
It moves the S2 beads in a mixed S1/S2 region to the band edges where they effectively control the material flow. Even when the band grows by incoming S1 grains, these are transported into the center of the stripe and the edge region remains occupied by the larger S2 grains. This is demonstrated with small colored (S2', 0.63-0.71 mm) and transparent (S1', 0.5-0.63 mm) spheres in Fig.~\ref{Fig3}, right. Initially, two bands were prepared where one contained only S1', the other one contained only the S2' fraction. The S1' material is transported into the growing stripe and in there, microsegregation places the S1' fraction in the stripe center. The edges between both S1' and S2' bands are astonishingly sharp, even though both species are neighboring sieving fractions.
This result agrees qualitatively with the segregation experiments by Newey {\it et al.} \cite{Newey04}
where they used three bead types with size ratios 0.6:1:2 and found that the bands of
large and small grains were separated by bands of the medium sized grains.

Most important, however, microsegregation does also occur in stripes of particles with a much narrower size distribution.
This can be verified by X-ray tomography in a down-sized system (tube diameter 24 mm, length 60 mm) filled
with beads of diameters $1.01\pm 0.01~$mm and 423 $\pm$ 23 $\mu$m
(roughly Gaussian diameter distribution, measured with a Retsch Camsizer).
Measurements were
performed in a Nanotom (GE Sensing and Inspection)
with 40~$\mu$m voxel size~\footnote{A voxel is the three-dimensional equivalent of a pixel.}.
The appendix describes in detail how  tomographic images of this resolution can be used to determine
the average particle diameter $d_{avg}$ with an accuracy better than a $\mu$m.

\begin{figure}[htbp]
	\begin{center}
	\includegraphics[width=0.6\columnwidth]{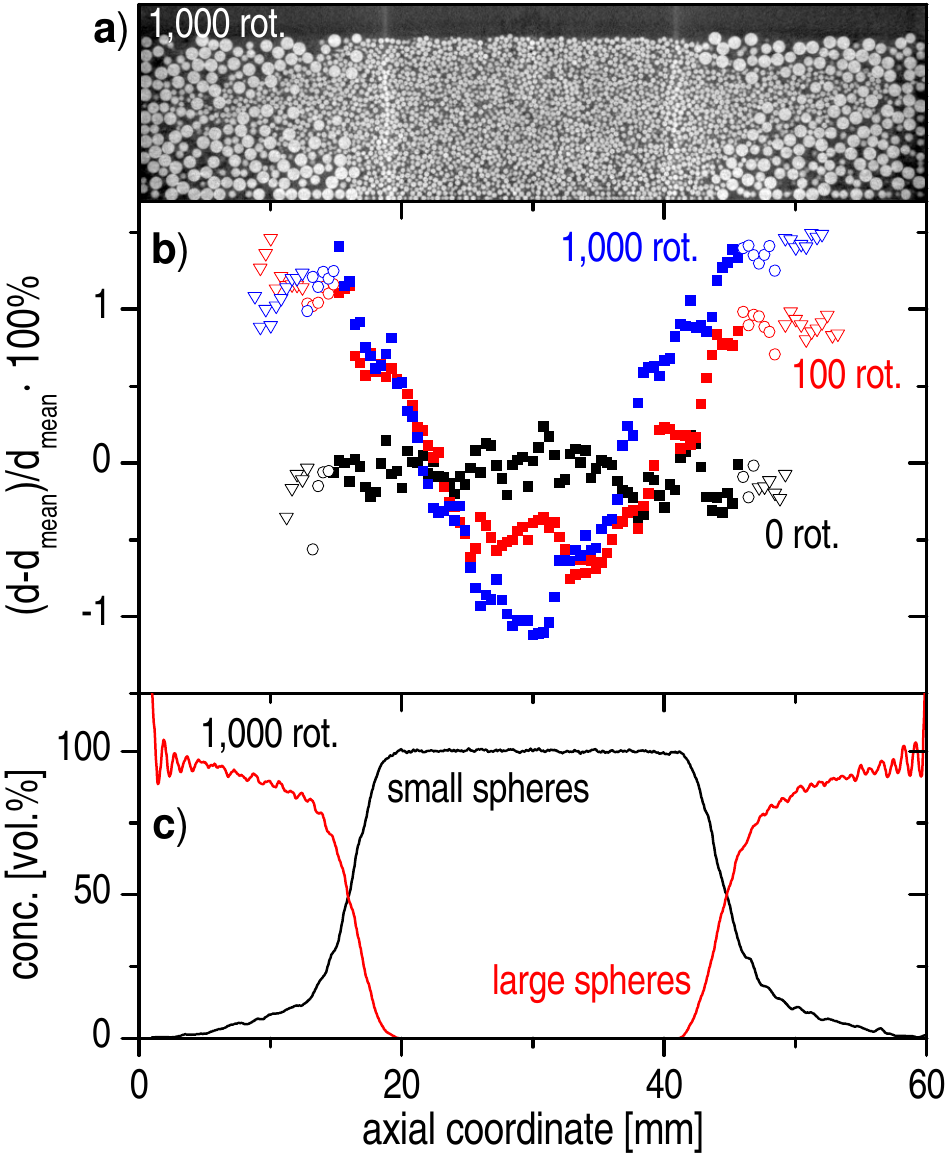}
	\end{center}
	\caption{Axial microsegregation. {\bf a)} Cross section, parallel to the tube axis, from a tomogram taken after 1000 rotations. {\bf b)} Axial microsegregation inside a single stripe of small beads (423 $\pm$ 23 $\mu$m)
surrounded by larger beads after 0, 100 and 1000 rotations. The average diameter in dependence on the axial positions is measured  using x-ray tomograms. Symbols indicate a bin size of 1.64~mm ({\tiny$\blacksquare$}),  4.04~mm ($\circ$), and 8.44~mm ($\triangledown$).
There is a clear tendency for the larger grains to accumulate at the band edges. {\bf c)} Concentration profiles along the tube axis for small and large beads after 1000 rotations.
}
\label{fig:microsegregation}
\end{figure}

After detection of the individual small and large particles we measure their average diameter
as a function of the axial coordinate.
The analysis of the small particle diameter in
homogeneously prepared single stripes of small grains shows
that beads are redistributed and microsegregated by the rotation of the mixer.
Figure \ref{fig:microsegregation}b shows the axial profile of the average diameter of the small grains in such a stripe.
Initially, the mixture is uniform (black, horizontal curve).
After 1,000 rotations of the mixer, a stable microsegregation is established
and beads near the edges of the stripe are on average 2~\% larger than those in the stripe center.
As demonstrated below in Sec.~\ref{sec:radseg}, we observe radial microsegregation as well. However,
this effect is not related to the flow through the core channel, thus we focus here on the axial microsegregation only.

\subsection{Particle size determines direction of long-term coarsening}

\begin{figure}[htbp]
	\begin{center}
	 \includegraphics[width=0.6\columnwidth]{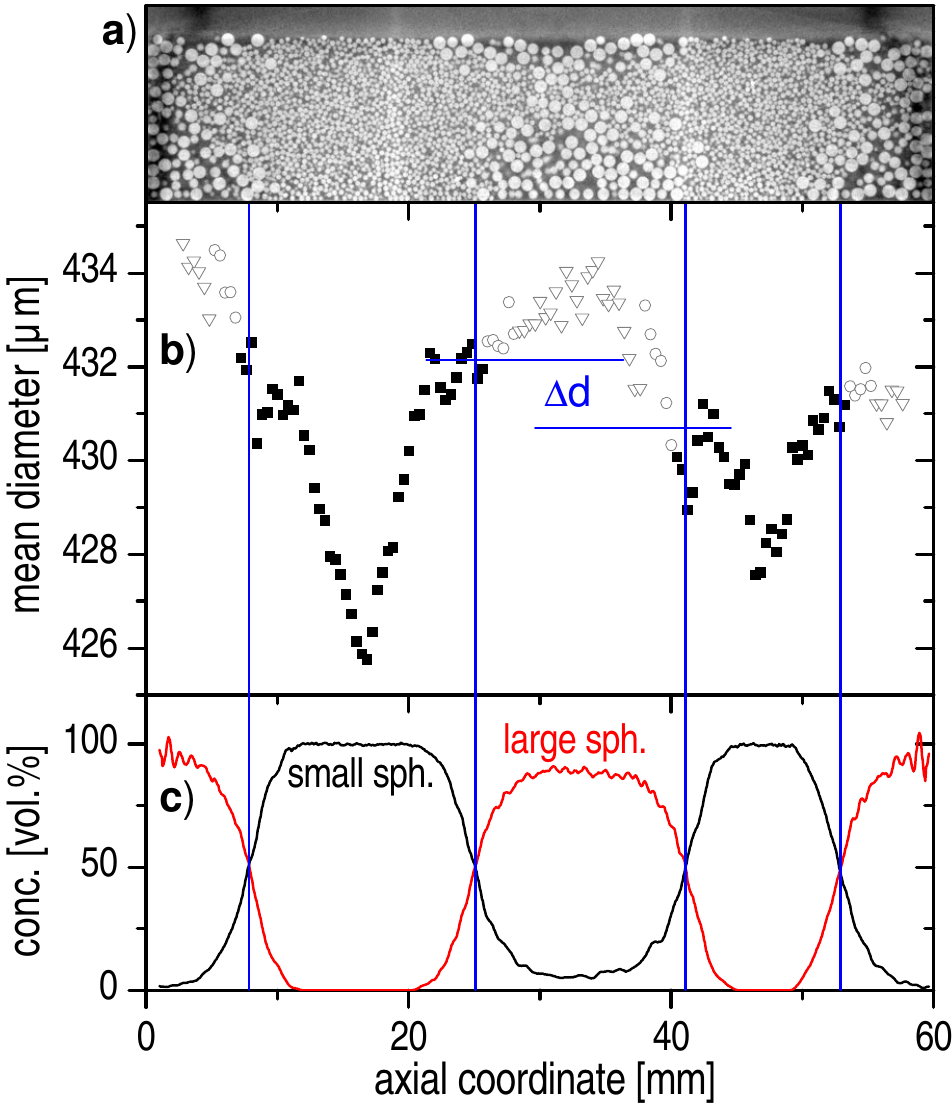}
	\end{center}
	\caption{Microsegregation drives the coarsening. {\bf a)} Cross section, parallel to the tube axis, taken from a tomogram
          {\bf b)} Microsegregation introduces an effective difference in bead diameter at the edges of
          adjacent bands of formally identical bead species; here 423 $\pm$ 23 $\mu$m.
          All data are measured using X-ray tomography.
          Blue vertical lines mark the stripe boundaries (equal volumes of small and large spheres).
          Symbols indicate the bin size used for the diameter measurements:
          1.64~mm ({\tiny$\blacksquare$}), 4.04~mm ($\circ$), and 8.44~mm ($\triangledown$)
          {\bf c)} Concentration profiles along the tube axis for small and large beads.
          }
\label{fig:sketch}
\end{figure}

It is this axial microsegregation that is responsible also for
the extinction of the narrower of two neighboring bands of beads with nominally the same diameters (c.f.~figure \ref{Fig1} b).
This is demonstrated with the experiment shown in figure \ref{fig:sketch}.
Using again X-ray tomography, we show that in a rotating drum prepared with two stripes
of the same small particles, two combined features occurred: the narrower stripe dissolves {\it and}
the dissolving stripe displays a smaller mean bead diameter at its boundaries.
This confirms once more that the direction of the flux between is pointing from the stripe containing smaller particles
to the one with larger particles.

Two considerations complete this argument. First, the smaller mean particle diameter at the
boundary of the narrower stripe can be understood as a consequence of the smaller reservoir inside its bulk which
can not provide the same amount of largest particles as they are present in the broader stripe.
Second, this mechanism is self-sustaining. Once the flow from the narrower
stripe to the broader sets in, the narrower stripe preferentially looses its largest particles which had accumulated at the boundary.

\begin{figure}[htbp]
	\begin{center}
	 \includegraphics[width=0.88\columnwidth]{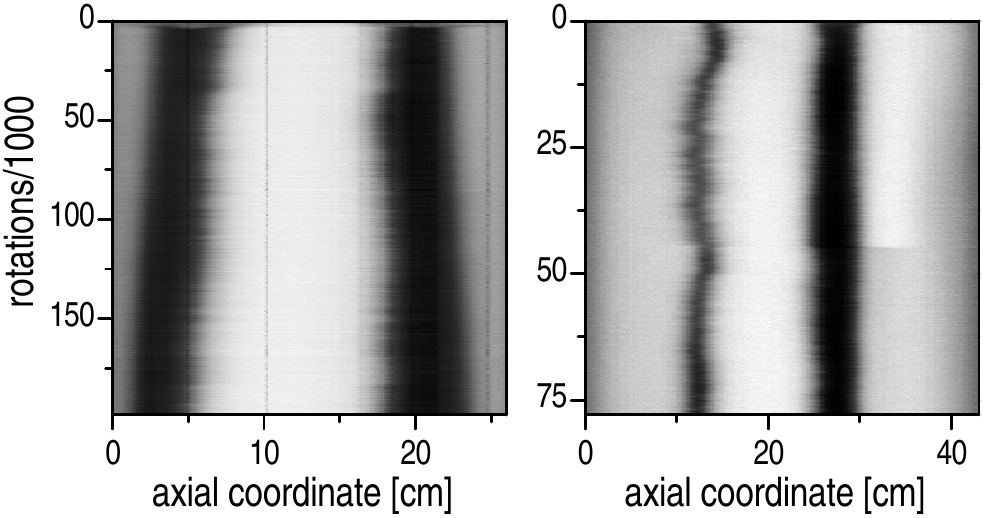}
	\end{center}
	\caption{Long-term coarsening is absent when the small particles are monodisperse. Systems were prepared with two small bands
composed of monodisperse small glass spheres (1.01 $\pm$ 0.01~mm) surrounded by larger (2.62 $\pm$ 0.065 mm) glass spheres.
Left: Space-time diagram of the evolution of two stripes of equal initial size.  Only drift but no coarsening is observed.
Right: Two bands prepared with  different initial widths, 5 cm and 8 cm, resp. Again no coarsening
 (but a pronounced lateral jitter) was observed. The tube diameter is 54 mm and the rotation speed is 20 rpm. }
\label{Fig7}
\end{figure}

If the above interpretation is correct, then the coarsening should be absent in systems where the small particles are monodisperse.
  Thus, we prepared drums with mixtures of large-size beads of 2.62 $\pm$ 0.065 mm and monodisperse small spheres of diameter 1.01 $\pm 0.01$ mm.
As shown in Fig.~\ref{Fig7},
the initially prepared bands  remain either stable or start to drift or jitter.
However, there is no coarsening on the timescale of the experiments, an order of magnitude longer than
the experiment displayed in Fig.~\ref{Fig1} b). The positional jitter of the thin stripe in Fig.~\ref{Fig7}b is an indication
that our experimental timescales are sufficient,
the slow exchange of large beads tunneling through the thin stripes becomes comparable to the small particle dynamics.

At this point the crucial question is, why the transport in the core
channel between two stripes is related to the particle size differences at the interfaces.
In Sec.~\ref{sec:hypo}, we propose two hypotheses:
the driving force could either be an effective surface tension which depends on the diameter ratio at the interface
between small and large particles. Alternatively, a gradient in small particle concentration  could determine the drift.
Either hypothesis requires further experimental verification.

Our findings agree well with observations of other researchers.
Rapaport \cite{rapaport:02,rapaport:07} reported numerical simulations
where the small (and also the large) beads had 20 \% size dispersion, and he
found a similar coarsening as in our experiments, after several thousand rotations.
Taberlet {\it et al.}~\cite{Taberlet04,taberlet:06_b} performed simulations with monodisperse beads and they found that a
steady state (without coarsening) was reached after a few hundred drum revolutions.
Newey et al.~\cite{Newey04} found zig-zag jitter of the segregated
bands without systematic coarsening in experiments with monodisperse beads, while experiments with slightly size dispersed small beads  \cite{Frette97,Fiedor03,Arndt05,Finger06,Finger07,Juarez08} clearly reproduce the long-term coarsening.
While all these earlier studies did not consider microsegregation, their results are in full
agreement with the conclusions of the present work.

\subsection{Radial microsegregation}
\label{sec:radseg}
\begin{figure}[htbp]
	\begin{center}
	 \includegraphics[width=0.6\textwidth]{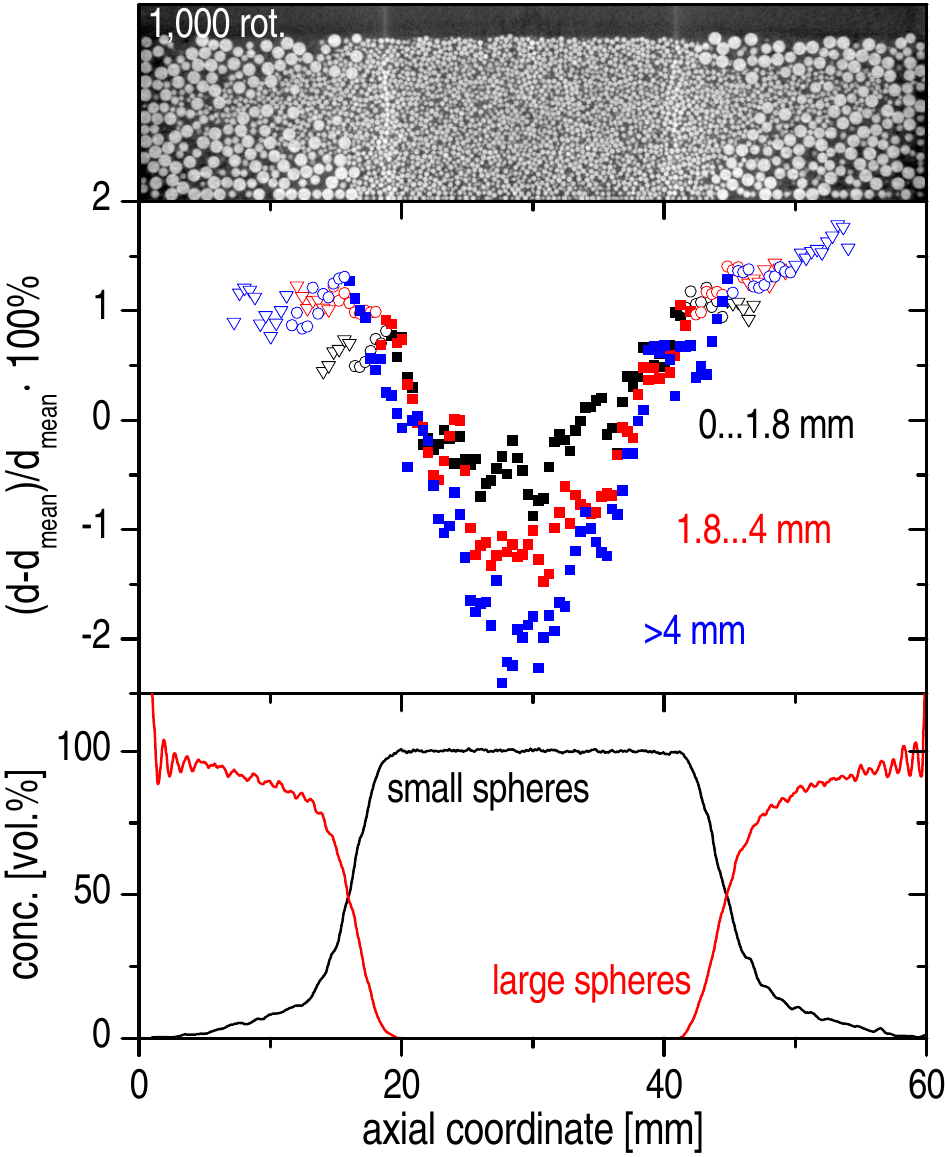}
	\end{center}
	\caption{Radial microsegregation. This figure corresponds to figure 3, showing data taken
after 1000 rotations. The three colors describe three radial bins where the distance is measured from the outer surface
of the sample. The blue data marked as $>$ 4 mm corresponds to the inner core of the sample.
Symoles indicate axial bin sizes of of 1.64~mm ({\tiny$\blacksquare$}),  4.04~mm ($\circ$), and 8.44~mm ($\triangledown$).
Bins contain at least 2000 (0 - 1.8 mm and 1.8-4 mm), respectively 1000 ($>$ 4mm) small spheres.
          }
\label{fig:radial_mcrosegregation}
\end{figure}

Microsegregation is also present in radial direction. Figure \ref{fig:radial_mcrosegregation} shows that the
mechanism establishes, especially in the center of the stripe, a gradient from smaller to larger diameters when moving
outwards from the center.
However, as this effect does not break axial symmetry it is not related to the uni-directional flow
behind the long-term coarsening phenomenon.

\section{Potential mechanism driving the flow during coarsening}
\label{sec:hypo}
One possible explanation is the existence of an effective surface tension $\sigma$ at the interface of each stripe
where $\sigma$ increases with the difference between small and large particle diameters $d_l - d_s$ \cite{schroter:12}.
It is always the stripe with the larger $(d_l - d_s)$ and therefore the larger  $\sigma$
which dissolves.  In order to explain the sequence of stripes evolving in Fig.~\ref{Fig3}, we need
to assume that $\sigma$ depends on $(d_l - d_s)^\alpha$  with an exponent $\alpha$ larger one.
An argument for the origin of $\sigma$ can be derived from
the assumption of an Edwards ensemble \cite{Edwards1989,Mehta1989,srebro:03,pica-ciamarra:12,bi:15}~\footnote{As the interparticle interactions (Hertzian normal forces and frictional tangential forces) do not display a strong size dependence, such a free surface energy will need to originate from entropic contributions.} where a configurational entropy $S_{\rm conf}$ is defined as the logarithm of the
number of possible mechanically stable configurations for the given volume fraction and boundary stresses.
The important point is that the number of possible packings of monodisperse
spheres does not depend on their diameter. Therefore in the bulk of both large and small stripes,
the entropy per particle $S_{\rm conf}/N $ is approximately identical.
 However, at an interface between two sizes of particles, the number of mechanically stable configurations will
depend on the diameter ratio. This could provide an effective surface tension $\sigma(d_l - d_s)$.
However, our experiment cannot provide direct evidence for this.

The second hypothesis presupposes that the linking channel contains a
concentration gradient  $dc_s/dx$ of small beads
that drives their directed motion.
We have established experimentally by varying the size ratios of large and small beads that the equilibrium concentration of
the smaller species in the core channel, $c_s$ (number of particles in the cross section) grows with the size ratio $r=d_l/d_s$. Different $r$ at the interfaces of two neighboring bands of small beads (caused by different compositions of the stripe edges) will therefore lead to a concentration gradient, $c_s$ becomes larger at the end where $r$ is larger (narrower segregation band or S1 beads). An effective particle flow towards the opposite channel end (large segregation band or S2 beads) is the consequence, until the band which provides the particles is completely extinguished.

None of these two hypotheses can be directly verified with the experiments presented here, and it is beyond the
goals of this study to prove the underlying physical origin of the observed mechanisms. Detailed measurements
of individual grain dynamics will be necessary for a validation, and a different experimental technique is
required for that.

\section{Summary}
Summarizing, two mechanisms are revealed here that govern the coarsening of axially banded pattern of bidisperse granular
mixtures in a horizontal rotating drum: microsegregation accumulates
larger particles of the small component at the band edges, and material transport generally sets in from bands
with smaller particles at their edges to neighbors containing larger particles at the edge. When the small-sized
beads are monodisperse, no coarsening of the pattern occurs, even over long experimental timescales.
We demonstrated that the stripe stability can thus be controlled by addition of a few percent of different-sized grains.
Furthermore, the observed microsegregation has been described here for the first time. It might be exploitable in technological processes where particles are to be sorted efficiently by size.

\section*{Appendix: Data processing of x-ray tomograms}

\begin{figure}[htbp]
	\begin{center}
	 \includegraphics[width=0.6\textwidth]{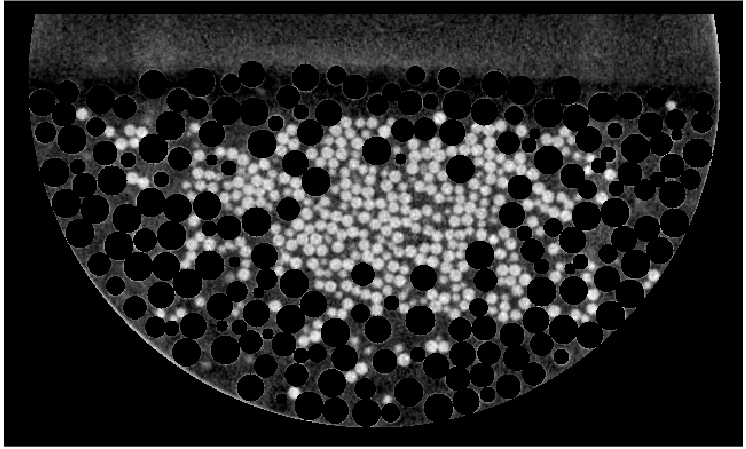}
	\end{center}
	\caption{Cross section through a tomography demonstrating the initial step of the image processing:
         the masking of the identified large particles.
}
\label{fig:detect}
\end{figure}

The initial step is a {\it precise determination of the centers of all small and large particles}. This task is performed with
a suite of matlab programs which first determine the centers of all large particles by finding the maxima in a
convolution of the tomography with a template of a large spheres. Then all large particles are removed from the tomography
by setting the corresponding voxels to black (c.f. Fig.~\ref{fig:detect}).

In the next step all small spheres are detected with again
voxel resolution by a second convolution with a small sphere template.
Finally, the position of the small particles is measured with sub-voxel accuracy
using an interpolation of the convolution results in the direct neighborhood of the maximum.

\begin{figure}[htbp]
	\begin{center}
	 \includegraphics[width=0.48\textwidth]{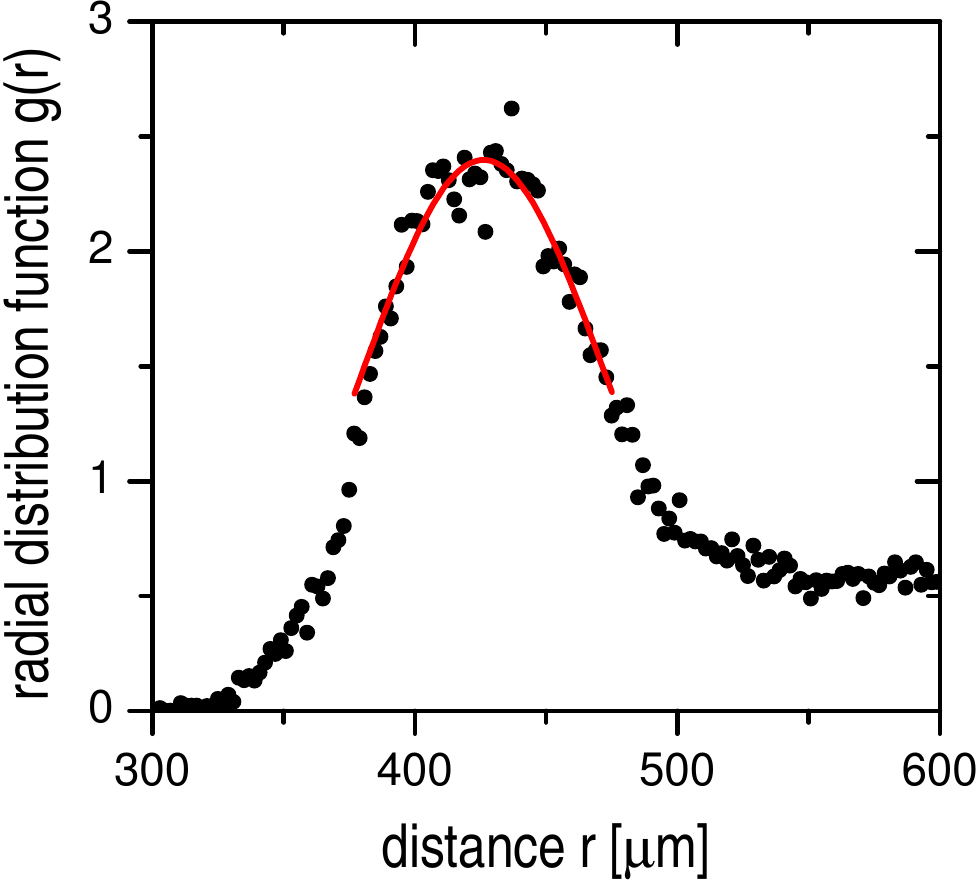}
	 \includegraphics[width=0.48\textwidth]{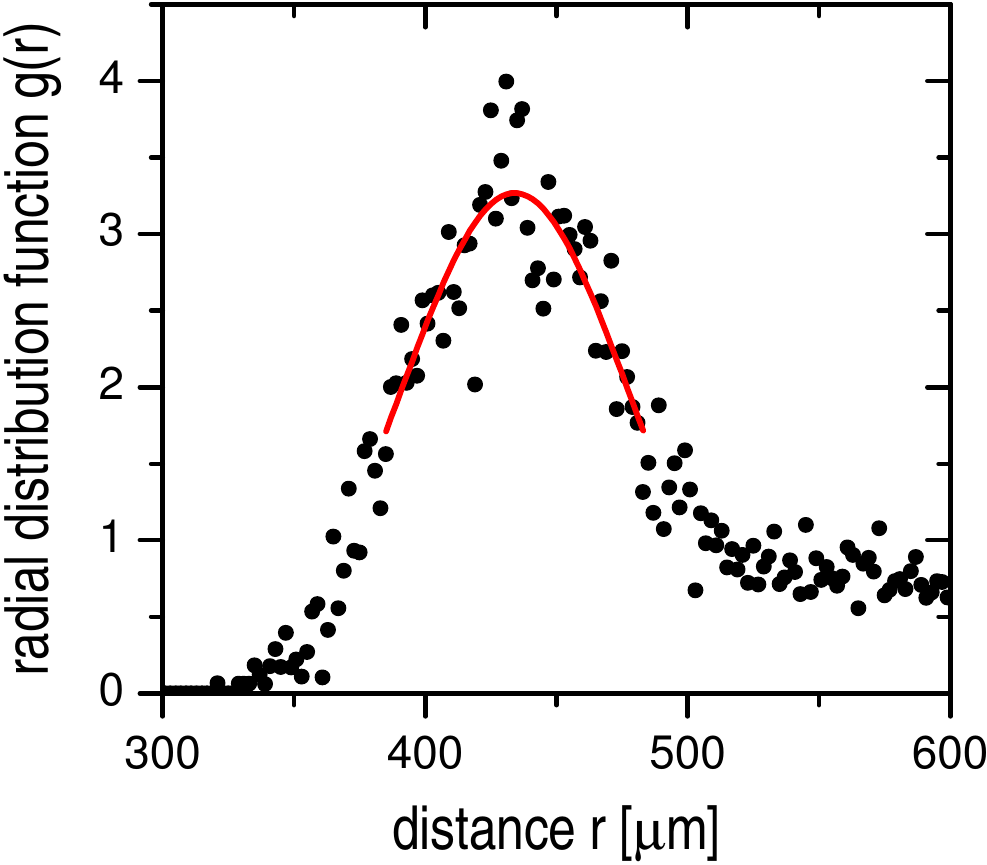}
	\end{center}
	\caption{The radial distribution functions computed for different slices with minimum and maximum
numbers of spheres in Fig.~\ref{fig:microsegregation}. Left: slice of width  $w = 1.64$~mm
at the axial position of $x = 16$~mm. This slice contained 5606 small spheres.
Right: slice of width  $w = 8.44$~mm at the axial position of $x = 32$~mm.
The red lines are fits to the peaks using Eq.~(\ref{eq1}.
  }
\label{fig:RDF}
\end{figure}

The second step is the {\it computation  of radial distribution function $g(r)$} which gives the probability to find the
center of a neighboring sphere within a shell of radius $[r, r+\Delta r)$. In our analysis we approximate $g(r)$ by using
shells with a width  $\Delta r$ = 2 $\mu$m i.e.~0.05 voxel. Figure \ref{fig:RDF} gives two examples.

The final step is based on the fact that the first peak of the radial distribution function,
i.e.~the most likely distance between two particles, corresponds to them touching.
Therefore {\it determining the position of the first peak of $g(r)$ with a Gaussian fit} is also a way to measure
the average diameter
$d_{\rm avg}$ of all particles. We perform these fits with
\begin{equation}
g(r) = a \cdot \exp\left({- \frac{(r - d_{\rm avg})^2}{2 \sigma^2}}\right)
\label{eq1}
\end{equation}
in an interval of $\pm 0.05~$mm around the approximate maximum determined by low-pass filtering of $g(r)$.

In Figs.~\ref{fig:microsegregation} and \ref{fig:sketch} we present the evolution of $d_{\rm avg}$ along the center $x$-axis of the cylinder
by computing $g(r)$ for all small spheres within bins of size $x \pm w/2$. However, in order
to have a sufficient statistic we have to adapt the window width $w$ to guarantee that there are at least 2000 small
particles within each window. The two graphs in Fig. \ref{fig:RDF} correspond to bins with the minimal and maximal number of small spheres included in the computation of $g(r)$.

\section*{References}
\def\newblock{}

\end{document}